\documentclass[letterpaper, 10 pt, conference, final]{ieeeconf}  

\IEEEoverridecommandlockouts                              

\usepackage{tabularx}
\usepackage{subfig}
\usepackage{orcidlink}

\usepackage{utfsym}
\usepackage{graphicx}
\usepackage{multirow}
\usepackage{makecell}
\usepackage{booktabs}

\usepackage{array}

\newcommand\copyrighttext{%
    \footnotesize \copyright{ }2024 IEEE. Personal use of this material is permitted. Permission from IEEE must be obtained for all other uses, in any current or future media, including reprinting/republishing this material for advertising or promotional purposes, creating new collective works, for resale or redistribution to servers or lists, or reuse of any copyrighted component of this work in other works.}
\newcommand\copyrightnotice{%
    \begin{tikzpicture}[remember picture,overlay]
    \node[anchor=south,yshift=15pt,xshift=0pt] at (current page.south) {\parbox{\dimexpr\textwidth-\fboxsep-\fboxrule\relax}{\copyrighttext}};
    \end{tikzpicture}%
}

\usepackage{hyperref}

    \newcommand\countbasescenarios{59,253}
    
    \newcommand\countgdeaevents{3,566,864}
    \newcommand\countenvelopes{7,538}

\title{\LARGE \bf
scenario.center: Methods from Real-world Data to a Scenario Database
}

\author{\href{https://orcid.org/0000-0003-2339-8157}{Michael Schuldes}\orcidlink{0000-0003-2339-8157}$^{1\dag}$, \href{https://orcid.org/0000-0003-4826-9706}{Christoph Glasmacher}\orcidlink{0000-0003-4826-9706}$^{1\dag}$,  Lutz Eckstein$^{1}$
\thanks{The work of this paper has been done in the context of the SUNRISE project which is co-funded by the European Commission’s Horizon Europe Research and Innovation Programme under grant agreement number $101069573$.
Views and opinions expressed, are those of the author(s) only and do not necessarily reflect those of the European Union or the European Climate, Infrastructure and Environment Executive Agency (CINEA). Neither the European Union nor the granting authority can be held responsible for them.}%
\thanks{Additionally, this research is funded by the VVM project research initiative, promoted by the German Federal Ministry for Economic Affairs and Climate Action (BMWK).}
\thanks{$^{\dag}$These authors contributed equally to this work and share first authorship.}
\thanks{$^{1}$Michael Schuldes, Christoph Glasmacher, and Lutz Eckstein are with Institute for Automotive Engineering RWTH Aachen University, Aachen, Germany {\tt\small firstname.lastname@ika.rwth-aachen.de}}%
}

\begin{document}

\maketitle
\thispagestyle{empty}
\copyrightnotice

\begin{abstract}

Scenario-based testing is a promising method to develop, verify and validate automated driving systems (ADS) since pure on-road testing seems inefficient for complex traffic environments.
A major challenge for this approach is the provision and management of a sufficient number of scenarios to test a system.
The provision, generation, and management of scenario at scale is investigated in current research.
This paper presents the scenario database \href{https://scenario.center}{scenario.center} to process and manage scenario data covering the needs of scenario-based testing approaches comprehensively and automatically.
Thereby, requirements for such databases are described.
Based on those, a four-step approach is proposed. 
Firstly, a common input format with defined quality requirements is defined. 
This is utilized for detecting events and base scenarios automatically. 
Furthermore, methods for searchability, evaluation of data quality and different scenario generation methods are proposed to allow a broad applicability serving different needs.
For evaluation, the methodology is compared to state-of-the-art scenario databases.
Finally, the application and capabilities of the database are shown by applying the methodology to the inD dataset.
A public demonstration of the database interface is provided at \href{https://scenario.center}{https://scenario.center}.

\end{abstract}

\section{Introduction}
Automated driving has great potential in increasing the efficiency and safety of traffic. 
To move more and more responsibility from humans to the driving function, the safety of such systems has to be established. 
Because of the open context \cite{sotif} of urban traffic, the development, and safety validation of driving functions through testing in real-world traffic is not feasible~\cite{Pegasus}.
To overcome this, scenario-based testing approaches have moved to the focus of research and industry to develop and argue for the safety of automated driving systems. 

The realization of scenario-based testing has a series of challenges.
To be able to define relevant test cases, an understanding of reality has to be established. 
This requires a large amount of diverse data from real traffic, which has to be processed and translated to scenarios. 
Because of the scale, such processes have to be automated and a managing and filtering of the resulting data has to be performed. 
This can be achieved through a scenario database.

Therefore, this paper proposes a methodology for a scenario database including the following steps:
\begin{itemize}
	\item Identification of scenarios in traffic data
	\item Computation of attributes and parameters of identified scenarios
	\item Storing and managing of scenario in databases
	\item Filtering of scenarios and providing data analysis inside the database
	\item Scenario generation for executable scenarios in simulation through OpenSCENARIO
\end{itemize}
The demonstrated methodology is implemented, and an application is shown. 
A publicly available, interactive demonstration of the scenario database is available under \href{https://scenario.center}{https://scenario.center}.

\section{Related Work}

In current research there are multiple fields approaching the topic of scenarios, as definition, processing, and storing of scenarios. For this paper, a scenario is defined as a sequence of scenes defined by actions and triggers \cite{RIE20}.

\subsection{Reality abstraction}
Despite different definitions, the goal of a scenario is the abstraction of reality to categorize complex traffic and simplify it for simulation. 
For this purpose, standards like OpenSCENARIO and OpenDRIVE are created as an interface between scenarios and simulation tools \cite{engel2020asam}. 
To systematically describe traffic and conceptually summarize properties, scenario concepts are defined \cite{WEB23} serving different abstraction layers as functional, logical and concrete scenarios \cite{Menzel2018}. 
In \cite{VVM_D13} this abstraction is refined by subdividing logical scenarios into logical scenario classes describing the parameter declarations and logical scenario instances including distributions. 
Orthogonal to these deviations, further categorizations are made.
\cite{SCH21} presents a basis for scenario concepts structuring scenario elements into 6 independent levels. These can be used to formulate an Operational Design Domain (ODD).
\cite{U_type} defines crash types focuses used to categorize describing accidents. These can be used to define a catalogue for scenario-based testing. 
\cite{Gelder2020} uses a tagging approach to describe scenarios comprehensively. 
\cite{WEB23} describes the actions and interactions of dynamic objects using a hierarchical structure based on concepts used to distiguish scenario categories in to base sceanrios/
Within \cite{VVM_D13} this concept is applied to more complex scenarios to describe recorded scenarios and generate scenarios for testing. Whereas recorded scenarios don't have a degree of freedom, this freedom is crucial in test scenarios to allow actions of a system under test.

\subsection{Processing of scenarios}
In principle, scenarios can be created in two ways, knowledge-based and data-based. 
The dividing line between those principles is fluid, going from fully knowledge-based scenarios, e.g., derived from ALKS requirements \cite{alks}, to fully data-based scenarios, e.g., data clustering, where even the scenario classes are derived from data, e.g., like performed in \cite{epple20}. 
\cite{9669219} demonstrates how scenarios according to the ALKS requirements are identified in data and derives concrete OpenSCENARIO files from that. 
\cite{Gelder2020} defines a rule based event identification from which scenarios are derived.
For data-based scenario extraction, working with different data sources is essential. 
This can be provided through the usage of formats like the OMEGA-format \cite{OMEGAformat}, as hown in \cite{VVM_D13}.

As an extension of identifying scenarios in data, one can generate new scenarios that fit to observed data. 
An example is \cite{VEHITS23}, where hybrid graphs are used to estimate the parameter distribution of observed scenarios and sample new, not seen scenarios from these distributions. 
An extensive meta overview of the scenario generation is given in \cite{schütt20231001}. 
The goal of the generated scenarios is to use them for testing and evaluation in simulation.
A widespread format, that is supported by various simulation tools is OpenSCENARIO~\cite{engel2020asam}.

\subsection{Scenario databases}
\label{sec:related_work_scenario_databases}
The idea of utilizing scenario databases for evaluating ADS was popularized by the Pegasus project \cite{Pegasus}. 
Within the project, highway traffic was captured and described. 
Pegasus structures traffic into a unified scenario concept, the challenger concept \cite{Weber2019}, enabling a unified parameter space covering the pre-defined highway scenarios.
But the description is limited to highway scenarios and safety relevant interaction with other traffic participants.
Several other approaches to manage scenarios efficiently exist. 
The scenario concept of Pegasus was developed further in \cite{Weber2023}, introducing the idea of base scenarios for the motorway context.

The Sakura scenario database \cite{scendb_sakura} spans from scenario identification to testing and evaluating test results. 
The scenarios are defined for highways. 
Starting from a test specification, corresponding predefined scenarios from a catalog are selected. 
For these scenarios, parameter distributions are extracted from data sources. 
These scenarios are translated to a test plan, which provides concrete scenarios to simulate.
The scenarios are described using OpenSCENARIO. 
The simulation results can be uploaded back into the database and evaluated. 
The available scenarios are derived from NCAP and ISO 34502, so deriving a relation to an ODD for coverage analysis is not directly given.

The ADScene database \cite{adscene} was developed out of the Moove \cite{brini:hal-04098011} project. As part of ADScene, a set of logical highway scenarios, together with their parameters is defined. 
Through a tool chain, concrete instances of these scenarios can be identified in data and aggregated to form parameter distributions for the logical scenarios. 
Automatic translation to simulation frameworks via formats like OpenSCENARIO is under development.

SafetyPool \cite{safetypool} can store functional and logical scenarios described in a specific Scenario Description Language (SDL), roughly comparable to OpenSCENARIO. 
Through either manually or automatically assigned tags, scenarios in the database are made searchable and the filtering to an ODD can be achieved.
A parameter distribution spanning more than one scenario of the same type (e.g., a joined parameter space for two different left turn logical scenarios) can not be achieved. The export of those scenarios to simulation tools via OpenSCENARIO is supported.

Existing scenario databases either lack the support of urban scenarios or are not able to create a unified parameter space over scenarios of the same type in the database. 
The latter is important when arguing for coverage and completeness. 
Additionally, current scenario databases do not support the formulation of filters for sequences of scenarios or events. 
The in this work presented \textit{scenario.center} aims at filling this gap.

\section{Methodology}
\label{sec:method}

This paper proposes a methodological setup of a scenario database, support urban use cases and a unified parameter space for the logical scenarios.
The complete process, spanning from automatic extraction of information from real-world data, over generation of logical scenario instances, querying, and generation of concrete scenarios is considered.

Before defining the structure, firstly, requirements are set for the database and underlying processes to serve different use cases for scenario-based testing:
\begin{itemize}
\item Scenarios have to be well-defined and understandable
\item Classification and processing into scenarios has to scale
\item Scenarios have to be understanble and searchable
\item Scenario generation and extraction should be efficient and flexible regarding the use case
\item Quality and coverage of the data should be estimated
\end{itemize}
In the following sections these requirements are used to develop methods which fulfill those requirements and enable efficient scenario-based testing.

\subsection{Database architecture}
\label{sec:architecture}

\begin{figure*}[t]
	\centering
	\vspace{0.035in}
	\includegraphics[width=\linewidth]{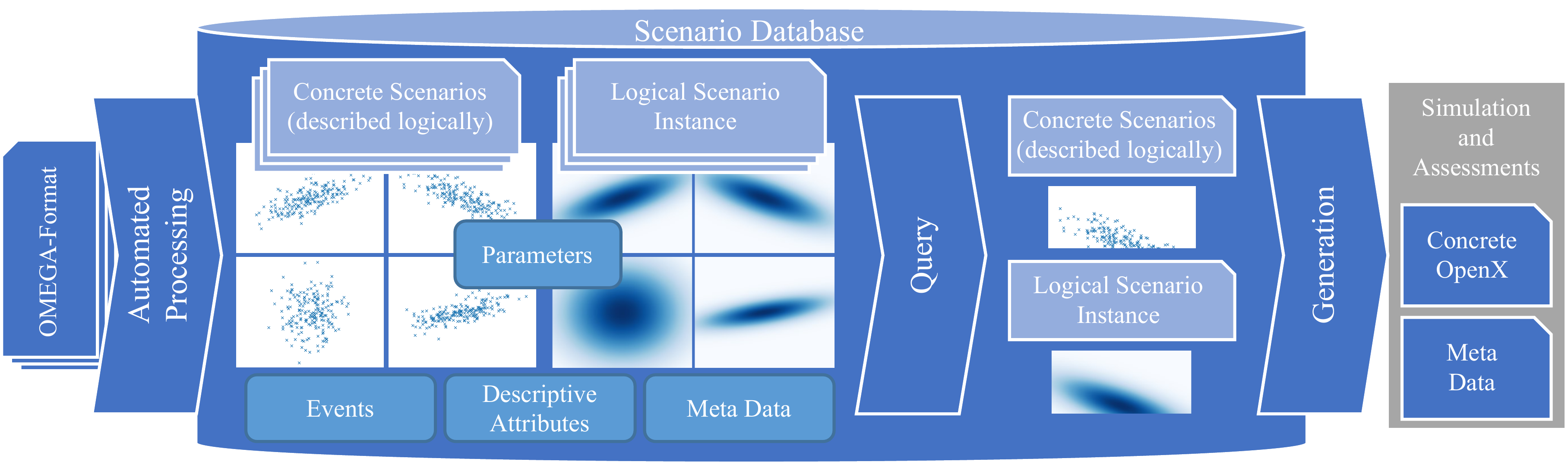}
	\caption{Scenario-database method, spanning from data acquisition to scenario generation.}
	\label{fig:database_method}
\end{figure*}

To allow an efficient setup and orchestration of different methods to serve the given requirements, firstly the structure of the database is discussed (see Fig.~\ref{fig:database_method}).

To ease the usage of different data sources and to provide a fixed interface to the database logic, an input format is defined. 
For the proposed method, the OMEGA-format \cite{OMEGAformat} is chosen. 
It provides converters for a diverse set of data sources and its definition follows the 6 Layer Model \cite{SCH21}. 

Based on that unified input, automated algorithms are applied (Sec.~\ref{sec:extraction}) to identify sequences or events in data according to an underlying and database overarching scenario concept (Sec.~\ref{sec:scenario}). 
The concept itself ensures a consistent representation of scenarios within the complete traffic domain.
For flexible usage, categorized data as well as extracted parameters are stored both as concrete instances and logical scenario instances with distributions or value ranges.
This combined storage allows the usage of real-world data as well as more sophisticated sampling methods for testing.

It might not always be of interest to work with a complete set or distributions of the whole available data. 
For example, only specific weather or types of intersections might be of interest. 
Therefore, a filtering step is included (Section \ref{sec:query}).
From there on, to derive test cases, the logical scenario instances or the sets of concrete scenarios can be sampled. 
However, viable sampling methods are not in the scope of this paper. 

After acquiring a concrete scenario to test or analyze, another key aspect of the scenario concept implemented in the scenario database comes into action: the ability to derive executable scenarios from each possible concrete instance of a logical scenario (Section \ref{sec:gen}). 
OpenSCENARIO as an output format of the scenario, since it is supported by many industry-standard simulation tools.

\subsection{Scenario concept}
\label{sec:scenario}

\begin{figure}[bt]
	\centering
	\includegraphics[width=\linewidth]{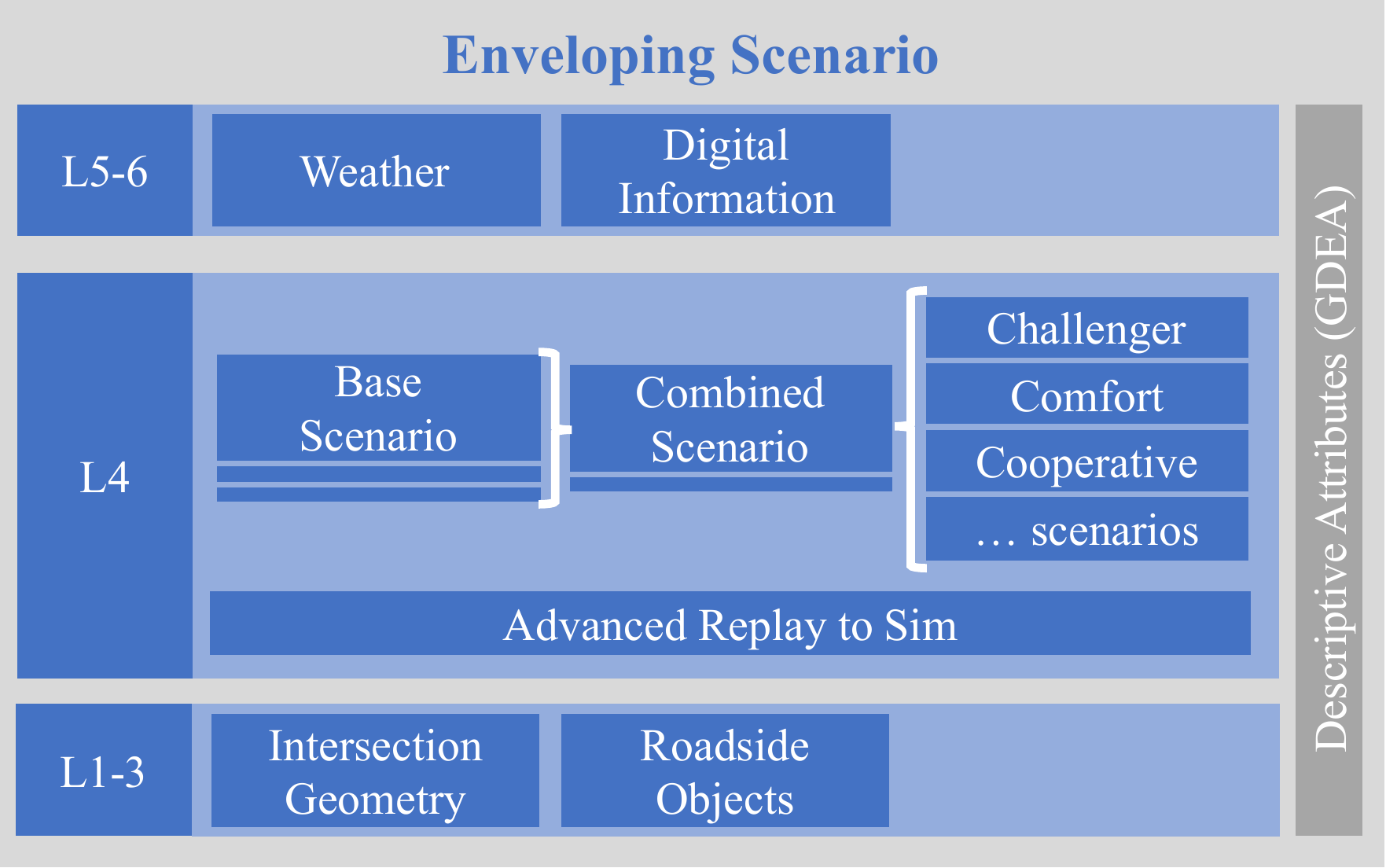}
	\caption{Database scenario concept}
	\label{fig:scenario_concept}
\end{figure}

In order to define scenarios for traffic unambiguously, a comprehensive scenario concept is used for classification (see Fig.~\ref{fig:scenario_concept}). 
The concept is thereby structured along the 6 Layer Model and utilizes the base scenario concept of \cite{WEB23}.
Longer traffic sequences are cut to enveloping scenarios, a spatial and temporal limited scenario with an assigned ego vehicle.

For this, Layer 1 ``Road Network and Traffic Guidance Objects'' and Layer 4 ``Dynamic Objects'' are used as primary structuring elements.  
Within the concept abstract scenarios are described as well as attributes and parameters to concretize all the layers presented. 
Attributes have a descriptive character and primarily serve the annotation and discoverability of scenarios. 
Parameters, on the other hand, serve to generate scenarios and should therefore be orthogonal to one another or at least have well-defined relations. 
Both are derived from applied concepts, which serve the description of the respective concepts for the creation of base scenarios. 
In this way, the hierarchical structure ensures consistency in that the core elements are described both via parameters and attributes.
Based on those, more complex combined scenarios can be created.
In addition to the parametrized approach, an unparameterized approach is used to allow more detailed and real-world based scenarios. 

\subsection{Automatic extraction of scenarios}
\label{sec:extraction}

To reach an adequate coverage of the ODD, a large amount of data has to be inserted into the database.
This requires a fully automated processing of data, that extracts scenarios and events.
In the following the such a method is described.
Each recorded drive, either real or simulated, is stored in the OMEGA-format. 
In a first step, the map information (Layer 1-3) is processed. 
Intersections and roundabouts are identified, and descriptions are derived.
From a calculated intersection center, the conflict area and the angle of incoming roads are computed. 
Additionally, semantic relations between lanes are determined. 
For the description to be compact and robust, the lane sections are automatically merged and split, to, for one, respect the calculated intersection area, and ensure consistency between variations in labeling.
With estimated lane and road centerlines, all objects are brought in association with lanes through 
Frenet coordinates similar to OpenSCENARIO.
These relative coordinates are essential for the scenarios and parameters to adapt to changing road networks. 

For the extraction of dynamic information, longer driving sequences are cut into enveloping scenarios in accordance with the scenario concept.
Enveloping scenarios take the perspective of an ego vehicle and split its drive into segments based on infrastructure characteristics. 
Within each enveloping scenario, various analyses are performed automatically. 
A distinction is made between event detection and the detection of base scenarios. 
Events assign individual attributes to specific points in time, e.g., the first point in time an object becomes visible for the ego. 
In contrast, base scenarios consider the semantic relationships between two road users over a time span.
To detect both types, expert-defined algorithms based on common metrics are used. 
Furthermore, additional metrics are defined or adapted to serve the detection of e.g., interactions in urban traffic.

\begin{figure}[tbh]
	\centering
	\includegraphics[width=\linewidth]{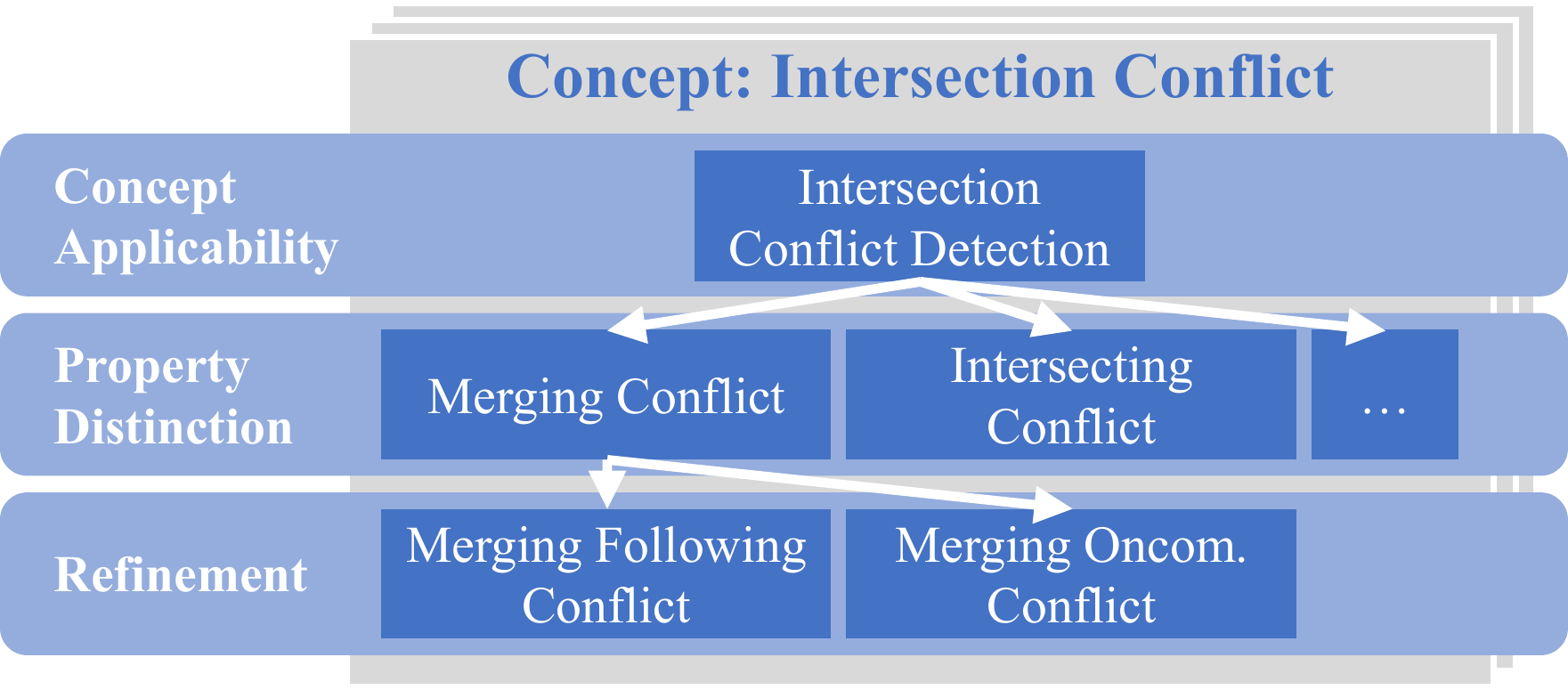}
	\caption{Hierarchical detection of scenarios}
	\label{fig:hierachical_detection}
\end{figure}

For efficiency, the process is designed hierarchically, so that metrics that are used multiple times only have to be calculated once and information can be detailed as needed and additional scenarios or events can be easily added (see Fig.~\ref{fig:hierachical_detection}).

\subsection{Querying scenarios}
\label{sec:query}

To navigate through a large number scenarios efficiently relevant information must be easily accessible and dedicated to the individual use case and needed scope. 
Since relevance of certain inforamtion can differ between users, a flexible filtering function is required.

The flexibility is reached by filtering for specific scenarios categories, individual attributes such as metrics, parameter ranges and events.
For efficient querying, this filtering options make use of the structure of processed data and the scenario concept.

The above-mentioned filtering options are commonly exposed in scenario databases. 
They allow the selection of specific scenarios but lack the possibility to define more complex relations and sequences of scenarios or events. 
To enable this, a graph-based definition of the query is developed.
Such user interfaces are known from rendering and video editing software like Blender~\cite{moioli2022introducing}. 
To allow an intuitive usage, users can graphically create nodes, change node features, and connect them up to form complex processes. 
This concept is adapted to define queries for complex scenarios and scenario sequences.
To define flexible queries with the tool, it is important to define fix interfaces between nodes.
Therefore, three types of nodes are defined: 
\begin{itemize}
    \item \textbf{Source nodes} specify road users and intersections
    \item \textbf{Filter nodes} specify filtering criteria as events and base scenarios
    \item \textbf{Result nodes} specify the output format and data for final or intermediate results.
\end{itemize}
With adding, configuring, and connecting nodes, the user can specify certain road users driving on specific intersections, experiencing defined events or base scenarios.
Through a sequence filter node, multiple event or scenario nodes can be brought into relation, e.g., the base scenario \textit{a} is happening right after base scenario \textit{b}.
Even complex queries can be defined with such graph-based definitions.

The challenge of this approach is the translation of this query graph into a database query, such as an SQL-statement.
Each node has to be translated to a part of a query.
This is achieved through the use and composition of common table expressions (CTEs).
Each node can be represented through a CTE. 
Each node CTE has to have a fixed set of columns, so the filter nodes can rely on their presence. 
Each node takes the input CTEs and produces based on its filters a new output CTE.
Result nodes can produce an output for the user from a CTE, e.g., displaying the set of relevant concrete scenarios or a plot of a distribution. 

\subsection{Generation of scenarios}
\label{sec:gen}

Users of the database are likely to have different requirements and applications for scenarios.
While some may only want to use them to view and analyze traffic, others may focus on re-simulating existing scenarios or sampling scenarios based on distributions. 
These requirements cannot be met by a single comprehensive parameterization and generation. Therefore, different generation options are offered and developed in the context of the database: 
\begin{itemize}
    \item Replay to Sim
    \item Advanced Replay to Sim
    \item Abstracted and parametrized scenarios
\end{itemize}
Within the \textit{Replay to Sim} (RtS) approach, recorded scenarios are run according to the originally recorded scenarios. 
The recorded traffic events are exactly reproduced as specified in the database input, the captured data.
However, this can lead to unrealistic behavior, when the behavior of the ego object in simulation deviates from the observed behavior in the source data. 
E.g., if an ego vehicle breaks despite not breaking in the original data, collisions with rear vehicles can occur, which have not happened in original data. 

For this purpose, an \textit{Advanced Replay to Sim} (ARtS) approach is pursued, which overwrites the original trajectory of other road users in case of such conflicts by means of a driver model \cite{VVM_D13}. 
Furthermore, a driver model to control the speed is implemented in the OpenSCENARIO standard, which allows deviation from the original speed profile to a necessary extent, based on time to X metrics, such as time to collision (TTC) and time headway (THW), while maintaining the original path.

In addition to the methods based on the reproduction or modification of input data, a parametrized approach to model the scenario space and provied appropriate abstractions is used.
For this, the parameterization of the scenario concept is utilized.
As described in Sec.~\ref{sec:architecture}, input data can be described either as distributions or as concrete values. 
Concrete parameter values can be modified manually or can directly be translated into an OpenSCENARIO and OpenDRIVE file.
Another option is to sample concrete values from a given distribution.
For such sampling, a modular approach, presented in \cite{VEHITS23}, is utilized.
It describes relevant relations based on causal dependencies and mathematical constraints and incorporates those in the sampling logic to increase modeling performance.
Using this approach offers several advantages for a database: 
Dependencies are described sufficiently due to causal considerations but the explosion of parameter dependencies and resulting inaccuracies for describing all multivariate relations does not occur. 
Furthermore, using a modular approach based on elementary parts of the scenario concepts allows definitions with reasonable amount of work.
Sampling from these distributions gives concrete parameter value sets which are translated to executable scenarios.

\subsection{Quality of database}

Besides the analysis and generation of scenarios, another important part is to determine the quality of the database.
For this purpose, a distinction is made into four quality and robustness criteria:
\begin{itemize}
    \item Quality of the input data
    \item Well-definedness and completeness of the concept
    \item Quality of processing
    \item Coverage of scenarios in the database
\end{itemize}
Thereby, the importance and determination of these criteria is dependent on the specific use case. Following, those criteria, and the implementation of those are discussed.

A database is only as good as the input data is.
Therefore, minimum quality requirements for accuracy, object representation and predefined classifications are needed for input data. 
This necessary quality is defined in the OMEGA format to ensure this standard within data \cite{OMEGAformat}. 
Beside the input data, the quality of the underlying scenario concept is important to ensure the usability of the scenarios. 
Beside unambigiousness, understandability and the ability to create executable scenarios, a certain completeness for the concept is needed to ensure the correct classification and handling of all possible input data. 
For this purpose, the method for reasoning about completeness according to \cite{VVMCompleteness} is used. 
Although this is applicable for an individual use case, it cannot be done for a whole database without knowing the concrete application due to different requirements for the level of detail and described traffic. 
So, it is only approximable based on common use cases. 

Another quality factor for the database in the processing of data. 
According to Sec.~\ref{sec:extraction}, data is processed automatically without further labeling effort. 
As potential errors would have significant impact on the usability of the database, the quality of data processing within the database has to be ensured. 
Therefore, it has to be guaranteed that used algorithms are robust and tested against diverse input data. 
Furthermore, quality checks have to be performed frequently, especially when changing input data.

The fourth measure is the coverage of real-world traffic data. 
This quality measure is not directly related to the systematic of the database or quality of individual input data, but to the amount and diverseness of data in the database. 
Giving indications for this is essential for scenario-based testing to allow a proper argumentation.
To give indications about a certain coverage, a coverage approximation under input data biases can be used as described in \cite{CGL23coverage}.

\section{Application}

To prove the applicability of the proposed methods and architectures of the scenario database, they are implemented in the scenario database \textit{scenario.center}. 
As a data source, the urban trajectory dataset inD~\cite{inD} is used.
It contains 13.499 trajectories from four urban intersections with mixed traffic. 
This dataset is converted in the OMEGA-format and processed automatically by the proposed methods.
According to the scenario concept (see Sec.~\ref{sec:scenario}), base scenarios are derived from the given real-world data. 
Thereby, $\countenvelopes$ enveloping scenarios are detected. 
Within those, $\countgdeaevents$ events and $\countbasescenarios$ base scenarios are found. 
To assess the accuracy, 150 randomly chosen base scenarios are checked manually.
No errors, neither classification or nor timing related ones were found.
Based on those detected base scenarios, RtS, ARtS, and parametrized scenarios are generated. 
These scenarios are available as OpenSCENARIO and OpenDRIVE files, making them executable in a diverse set of simulation applications. 
Additionally, videos are created using esmini~\cite{esmini} and displayed in the user interface.

Besides the processing of individual scenarios, distributions are utilized within the database. 
Next to the generation of new scenarios as described in \cite{VEHITS23}, distributions are used and visualized for comparison of different data sources.
These distributions are the basis for later generation of scenarios. 

\subsection{Data analysis}
\label{sec:app_analysis}

\begin{figure}[tbh]
	\centering
	\includegraphics[width=\linewidth]{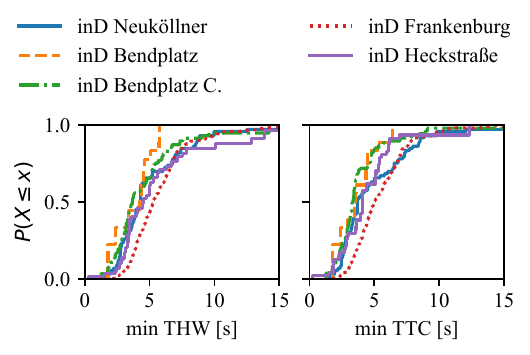}
	\caption{Parameter distribution (ECDF) of minimum THW and TTC of the scenario concept \textit{IntersectingConflict}.}
	\label{fig:parameters}
\end{figure}

Prior to generation, multiple analyses to compare and understand the underlying data can be performed with the database. 
This analysis step is key to have a robust simulation. 
The parameters and attributes of base scenarios can be explored in depth. 
Fig.~\ref{fig:parameters} shows empirical distribution plots of the minimum THW and TTC of the base scenarios with an intersecting conflict divided by the data source.
One can see that \textit{inD Frankenburg} has fewer scenarios with a low  TTC as well as fewer with a low THW compared to the other intersections. 
This can be explained through the difference in speed limit. 
The \textit{inD Frankenburg} has a speed limit of $30 km/h$, whereas the others have a speed limit of $50 km/h$.

\begin{figure}[tbh]
	\centering
	\includegraphics[width=\linewidth]{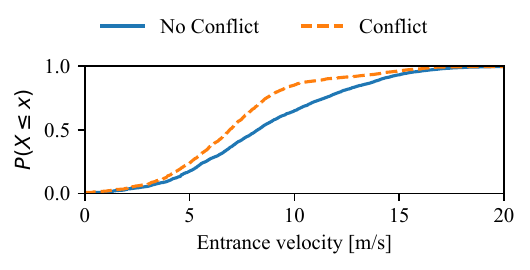}
	\caption{Entrance velocities of ego vehicles for crossings with and without a conflict scenario.}
	\label{fig:conflict}
\end{figure}

Not only the characteristics of different data sources and their influences on parameter distributions can be analyzed, but also the effects on object behavior of the scenarios themselves. 
As shown in Fig.~\ref{fig:conflict}, the intersection entrance speed of the ego vehicle tend to be lower when a conflict on the intersections occurs, compared to when no conflict occurs. 
A conflict thereby is a shared usage of same space within a predicated timespan. 
This does not mean, that approaching an intersection with higher speeds is safer, but that in most cases potential conflicts are identified in advance and the behavior is adapted to that conflict. 

\subsection{Data query}
\label{sec:app_query}

As indicated in Sec.~\ref{sec:query}, a user is only interested in a subset of scenarios at a time.
Either because of a reduced ODD or because a specific situation or specific circumstances need to be inspected in more detail. 

Designing an interface for filtering of metrics and attributes one to one related to the scenarios is straight forward. 
That is implemented with toggles for Boolean variables, (multi-) select-boxes for enumerations and sliders (and interval sliders) for scalar (interval) variables like float and times.

\begin{figure}[tbh]
	\centering
	\includegraphics[width=\linewidth]{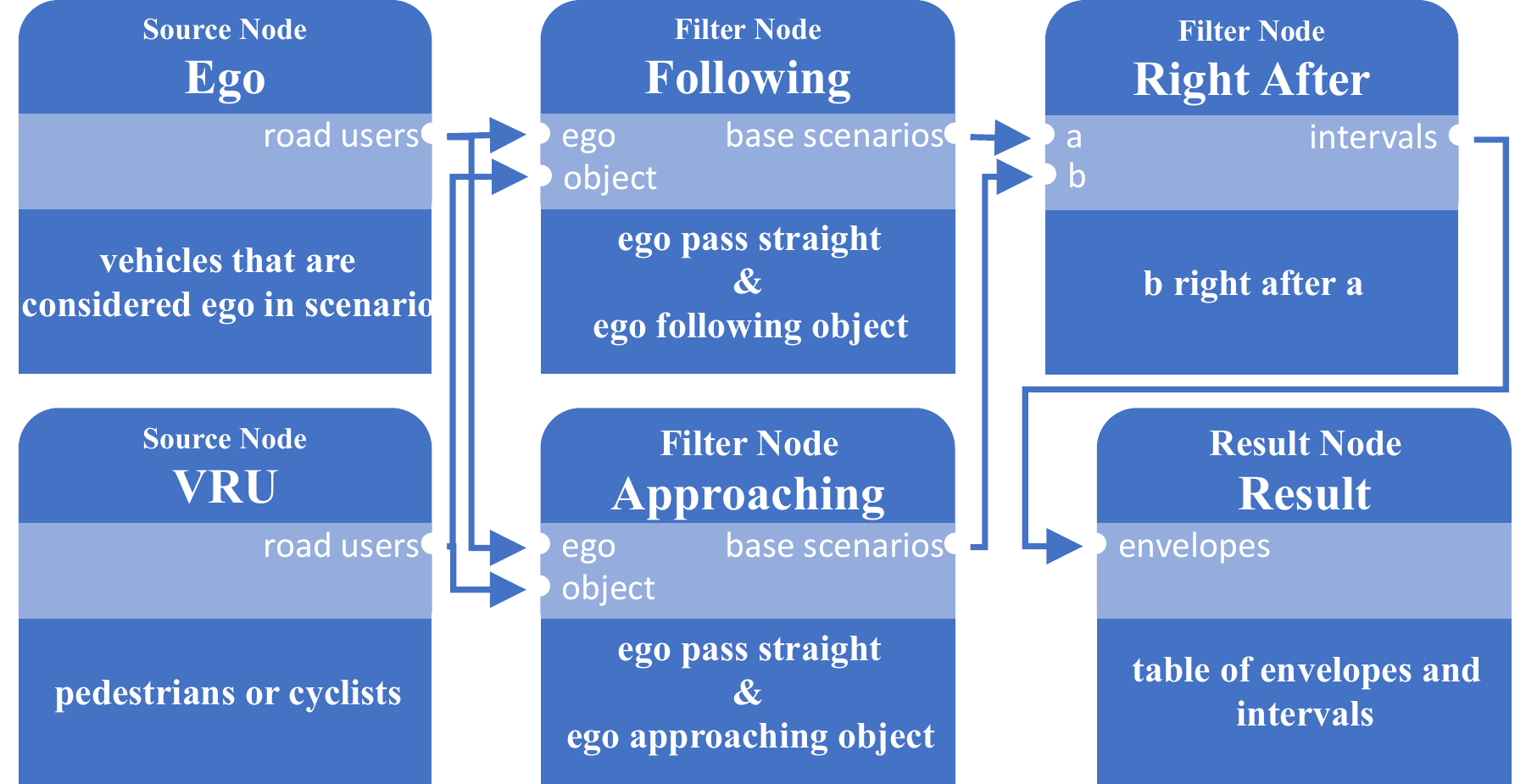}
	\caption{Abstraction of node based user interface to interactively create queries on sequences of scenarios and events.}
	\label{fig:sequence_query}
\end{figure}

As described in Sec.~\ref{sec:query}, the querying for sequences is designed in a node graph interface. 
Figure \ref{fig:sequence_query} shows an example of query formulated through the interface. 
Node \textit{Ego} and \textit{VRU} are source nodes and define desired objects, a vehicle regarded as the one experiencing the scenarios and a road user of type vulnerable road user, meaning a pedestrian or cyclist. 
More detailed filtering options in each node are possible, e.g., size of the object, maximum speed or color.
The filtering nodes, arbitrarily named \textit{Following} and \textit{Approaching} both define base scenarios where the ego vehicle moves straight through the intersection and is in following or approaching relation to the other object. 
Through the edge connections, it is clear, that the ego vehicle is \textit{Ego} and the object should be that from the \textit{VRU} node.
The third filtering node \textit{Right After} defines a time relation of both nodes. 
They should occur right after another. 
The last node defines the desired output format. 

\subsection{Display and generation}
\label{sec:app_display_generation}
\begin{figure}[tbh]
	\centering
	\includegraphics[width=\linewidth]{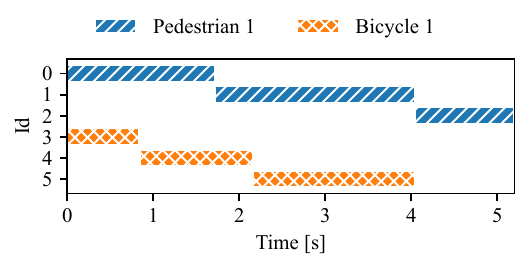}
	\caption{Scenario timeline of an enveloping scenario as a result from a query. The base scenarios are described in Table \ref{tab:sequence}.}
	\label{fig:sequence_scenario}
\end{figure}

One result of the query defined in \ref{sec:app_query} can be seen as a Gantt graph in Fig.~\ref{fig:sequence_scenario} and Tab.~\ref{tab:sequence}.
Each bar represents a base scenario the ego vehicle experiences with different road users and the table defines these base scenarios through specifying the concept values the base scenario is defined from. 
Scenario sequences found via queries, random samples or broader filters are visualized in different ways to allow an easy understanding of the situation.

\begin{table}[htp]
\centering
\vspace{0.053in}
\caption{Scenarios in sequence of Figure \ref{fig:sequence_scenario} (\textit{OTAC} = Object Traffic Area Change, \textit{ROP} = Relative Object Position, \textit{EM}= Ego Maneuver, \textit{LS} = Longitudinal State).}
\label{tab:sequence}
\begin{tabular}{|r||c|c|c|c|}
\hline
Id & OTAC~\cite{WEB23} & ROP & EM & LS \\
\hline
0 & - & oncoming & pass straight & - \\
1 & crossing & oncoming & pass straight & - \\
2 & crossing & - & - & - \\
3 & - & same arm & pass straight & following \\
4 & - & same arm & pass straight & approaching \\
5 & - & same arm & pass straight & - \\
\hline
\end{tabular}
\end{table}

\begin{figure}[tbh]
	\centering
	\includegraphics[width=\linewidth]{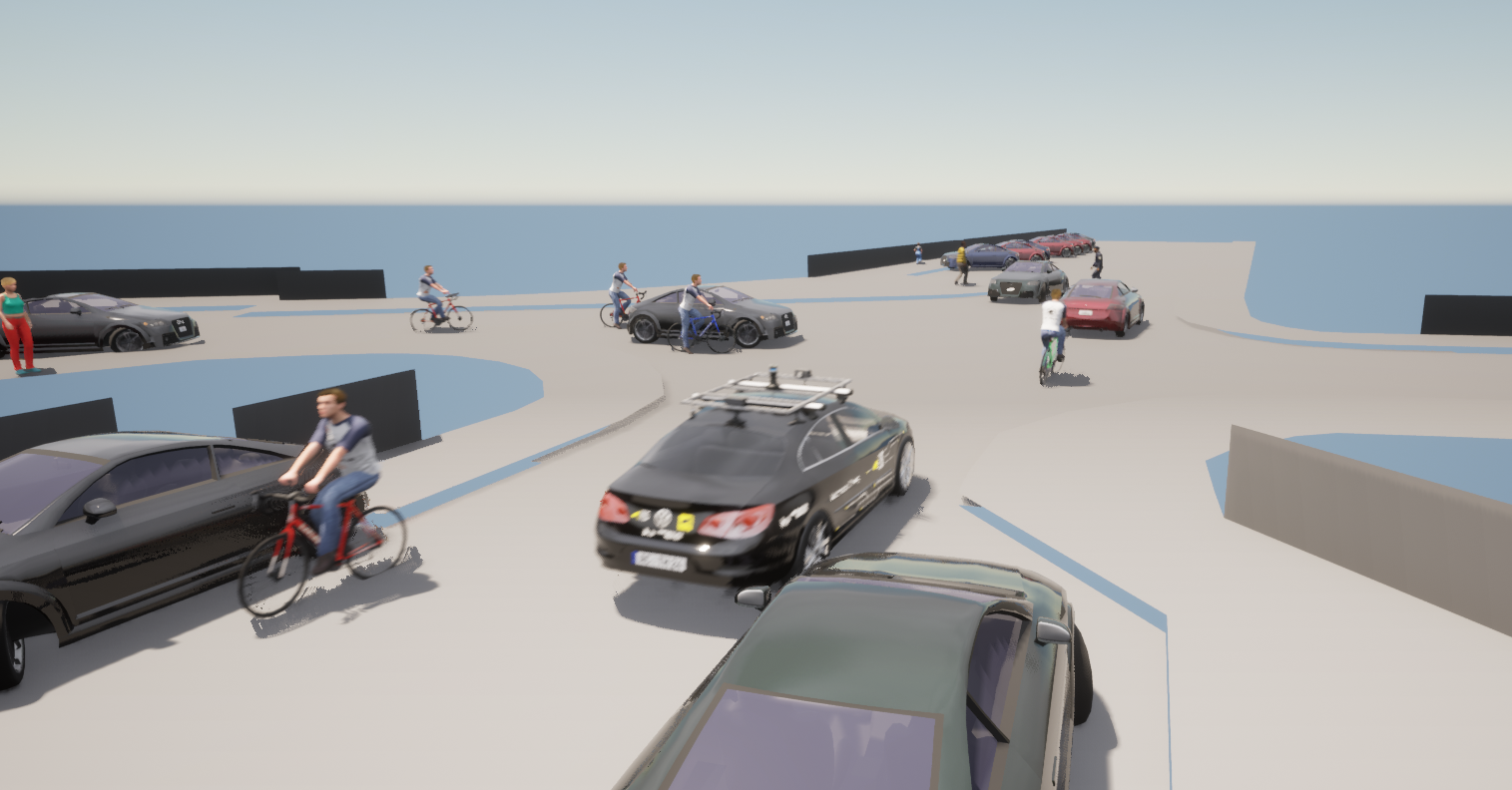}
	\caption{Example image of simulation result based on XOSC and XODR acquired from the database, produced with CARLA.}
	\label{fig:sequence_scenario_carla}
\end{figure}

Furthermore, a video is generated to visualize the scenario in form of RtS simulated with esmini.
Since the scenario is defined through OpenSCENARIO it can be simulated in CARLA, see Fig~\ref{fig:sequence_scenario_carla}. 
Next to those visualization tools, OpenDRIVE and OpenSCENARIO files can be downloaded to use files in own simulations.

Not only detailed information on individual found scenarios is of interest, but also the joined information from all found scenarios. 
Therefore, attribute and parameter distribution plots are available, comparable to those shown in Sec.~\ref{sec:app_analysis} Fig.~\ref{fig:parameters}, but only considering data from the selected scenarios.

\subsection{Test execution}

\begin{figure}[tbh]
    \centering	
    \includegraphics[width=\linewidth]{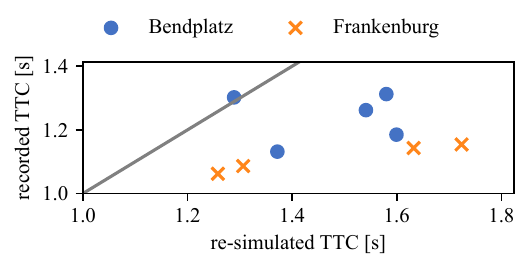}
    \caption{Comparison of recorded and re-simulated TTCs utilizing an ACC function. The identity line is marked in gray.}
    \label{fig:comparison_TTCs_CARLA}
\end{figure}

To check whether the generated scenarios from the database can be used to test driving functions, exemplary scenarios are taken from the query from Fig.~\ref{fig:sequence_query}.
These are used to test an automatic cruise control (ACC) function in CARLA. 
For this, the scenarios are automatically processed to a CARLA-readable format and for a simple demonstration the TTC is taken as a measure to assess the driving function.
Comparing the test scenarios with the real-world recorded constellations, lower TTCs in the simulation can be observed (see Fig.~\ref{fig:comparison_TTCs_CARLA}).
This suggests, that the example ACC drives more aggressive than the observed real-world drivers.

\section{Discussion}
This paper demonstrates the methodology of a scenario database designed for the urban use case, providing a unified parameter space for the extracted scenarios.
The demonstration of the methodology through application is successfully demonstrated and accessible publicly in \href{https://scenario.center}{scenario.center}. 
The whole process from identifying scenarios in data, over filtering to relevant scenarios, up to generating executable scenario in form of OpenSCENARIO was demonstrated. 

How the developed method compares to existing approaches of scenario databases introduced in Sec.~\ref{sec:related_work_scenario_databases} is shown in Table \ref{tab:scenario_database}. 
Only \textit{SafetyPool} and \textit{scenario.center} support the urban use case. 
The main differentiating factor is the underlying scenario concept.
It decomposes traffic into composable building blocks, which are applicable in all traffic situations. 
This is key to achieve a unified parameter space per type of scenario over all data sources and a comparison between data sources.
Only Pegasus demonstrated such a unified parameter space, though limited to the less complex highway setting. 
Such a unified parameter space also eases the argumentation for coverage and completeness, which is important for safety argumentation based on the database. 

Additionally, \textit{scenario.center} first demonstrates the application of a filtering interface for sequences of scenarios in a scenario database, and it is the only scenario database, which provides a publicly available demonstration version.
The latter is important to promote the understanding of the real world capabilities and use cases of the scenario database.

\begin{table}
\centering
\caption{Table comparing the different existing scenario databases.}
\begin{tabular}{|c||c|c|c|c|c|} 
\hline
Feature                           & \rotatebox[origin=c]{90}{Pegasus} & \rotatebox[origin=c]{90}{Sakura} & \rotatebox[origin=c]{90}{ADScene} & \rotatebox[origin=c]{90}{SaftyPool\textsuperscript{\texttrademark}} & \rotatebox[origin=c]{90}{ scenario.center }  \\ 
\hline
highway                           & \usym{1F5F8}                      & \usym{1F5F8}                     & \usym{1F5F8}                      & \usym{1F5F8}                          & \usym{1F5F8}                               \\
urban                             & \usym{2717}                       & \usym{2717}                      & \usym{2717}                       & \usym{1F5F8}                          & \usym{1F5F8}                               \\ 
\hline
automated extraction from data    & \usym{1F5F8}                      & \usym{1F5F8}                     & \usym{1F5F8}                      & \usym{1F5F8}                          & \usym{1F5F8}                               \\
unified scenario concept          & \usym{1F5F8}                      & \usym{2717}                      & \usym{2717}                       & \usym{2717}                           & \usym{1F5F8}                               \\
unified parameter space           & \usym{1F5F8}                      & \usym{2717}                      & \usym{2717}                       & \usym{2717}                           & \usym{1F5F8}                               \\
sample scenario from distribution & \usym{1F5F8}                      & \usym{1F5F8}                     & \usym{1F5F8}                      & \usym{1F5F8}                          & \usym{1F5F8}                               \\ 
\hline
filter by sequence                & \usym{2717}                       & \usym{2717}                      & \usym{2717}                       & \usym{2717}                           & \usym{1F5F8}                               \\ 
\hline
coverage \& completeness argumentation                       & \usym{2717}                       & \usym{2717}                      & \usym{2717}                       & \usym{2717}                           & \usym{1F5F8}                               \\
\hline
\end{tabular}
\label{tab:scenario_database}
\end{table}

Nevertheless, other frameworks provide benefits for different applications. 
\textit{SafetyPool} promotes the sharing of scenarios through incentives, which is beneficial for industry-wide cooperation and strengthening synergies. 
The Sakura database tightly integrates testing and test results into the user interface, which demonstrates the future of other databases. 
Therefore, to enable one user to use all the benefits at once, interoperability between the different solutions is desired. 
Such a federated layer, that connects scenario databases, is the focus of the SUNRISE project \cite{SUNRISE}. 

Although OpenSCENARIO is used as a standard to generate scenarios it has to be stated that the applicability has some limitations. 
On the one hand, many assessed frameworks only support a subset of the functionality of the standard and may use different interpretations. 
So, targeting a broad support of tools leads to a limitation of design opportunity of scenarios. 
On the other hand, whereas the functionality may be sufficient for highways, further triggers would be beneficial to generate and simulate scenarios on urban intersections, especially in lateral traffic movements. 
Within a common database a trade-off always has to be made between usability and functionality.

Lastly, for a scenario database to sufficiently reflect a desired ODD, incorporating as much data as possible is important. 
However, such data is only available in limited capacity, especially, publicly available data.
Moreover, the available data is of different quality and the used data formats vary, making incorporating lots of data difficult. 
In our work, we deal with the problem by utilizing the OMEGA-format~\cite{OMEGAformat} as an interface between data and scenario database. 
This eased the adoption of new data sources, highlighting the importance of agreed upon standards in industry and research.  

\section{Conclusion}

Within this paper, requirements and methods to create a scenario database are proposed. 
The developed methods are implemented in a new publicly available scenario database called \textit{\href{https://scenario.center}{scenario.center}}.

It is demonstrated how to automatically identify scenarios in data and compute corresponding parameters and attributes.
Furthermore, the management of scenarios at scale and the provision of insights and filtering options to users is shown.
Lastly, the generation of scenarios as input for simulation tools is described. 
These steps are demonstrated in an example, from acquiring data in a scenario database to assessing an example ACC.

Since there are multiple scenario databases with different capabilities and concepts, an orchestration of databases as discussed in the SUNRISE~project~\cite{SUNRISE} is future work.


\bibliographystyle{IEEEtran}
\bibliography{bibtex/bib/IEEEabrv, bibtex/bib/paper}

\end{document}